# Assessing the impact of cyber attacks manipulating distributed energy resources on power system operation

Philipp Linnartz, Alexander Winkens, Andreas Ulbig
Institute for High Voltage Equipment and Grids, Digitalization and Energy Economics (IAEW)
RWTH Aachen University
Aachen, Germany
p.linnartz@iaew.rwth-aachen.de

*Abstract*—Successful cyber attacks on power systems cause severe disruptions. One possible manipulation strategy is the utilization of distributed energy resources (DERs) to disturb power system operation. In addition to the impact on bulk power system frequency, local cascading effects caused by DER control and protection can increase the severity of this strategy. To investigate these effects, manipulation scenarios including the disconnection as well as the manipulation of active (P) and reactive power (Q) setpoints of DERs are derived. The impact is analyzed using time-domain simulations and quantified using assessment criteria such as voltage band violation and plant protection triggering. Though DER disconnection leads to high amounts of lost P injection the manipulation of Q setpoints offers potential to disconnect additional DERs through local cascading effects. To mitigate the impact of the manipulation scenarios automated tap changer operation as well as a limitation of remotely accessible Q is suitable.

*Index Terms*-- Cyberattacks, distributed energy resources, distribution grids, power system stability, time-domain simulation

I. INTRODUCTION

Nowadays information and communication technology (ICT) is deployed in the distribution grids (DGs) to utilize the flexibility of a rising number of distributed energy resources (DERs) for operational and market purposes. As a consequence, DERs are increasingly connected to the control and monitoring systems of the distribution grid operators (DSOs), virtual power plant operators (VPPOs) and/or the internet of things (IoT). A cyber-physical system (CPS) is thus emerging [1].

However, during the last decade CPSs (e.g., the electric power system) have been increasingly targeted by cyber attacks [2, 3]. A successful cyber attack on power system infrastructure can lead to severe service disruptions, economic losses and physical infrastructure damage [3]. The general vulnerability of the power system infrastructure and the severity of the impact of cyber attacks are shown, for example, by the successful cyber attacks on Ukrainian distribution system operators starting in late 2015. After gaining access to the central control system, several distribution substations were disconnected by the attackers leading to a local blackout [4].

Although the impact of this rather straightforward manipulation strategy opening the breakers at distribution substations is conclusive, also more subtle and harder to detect attack strategies targeting the electric power system can be envisioned. The aim of these attack strategies is to compromise availability, confidentiality, and integrity of the CPS [5]. The attack scenarios can be further classified into control-based and measurement-based attacks. The former can directly cause frequency and transient voltage instability, line overloading, load shedding, as well as cascading failures by manipulating or forging control signals sent to the targeted power system assets. The latter aims to compromise measurements to hide or falsify the current state of the system, weaken observability and eventually mislead operators or control systems [6]. A variety of hypothesized attack schemes targeting the generation, transmission, distribution, customer, and electricity markets domain exists and are investigated. Examples are: denial of service (DoS) attacks that flood and congest the network with maliciously generated traffic [7], data integrity (DI) attacks that manipulate control or measurement signals [8], and switching attacks where assets such as power lines, loads or DERs are switched on and off to cause instabilities and/or outages in the power system [9]. For a thorough analysis of their impact in the power system domain, the dynamics of power system assets, control and protection circuits need to be considered [10].

One particular attack scenario is the utilization of a large number of DERs connected to the DGs to disturb stable power system operation. This can be achieved by sending manipulated control signals. Due to their high penetration in the DGs, their control algorithms and protection functions have a crucial influence on that impact [10]. Current research analyzes the impact of this attack scenario by quantifying the impact using aggregated models of DERs or equivalent models of DGs

This work has been performed within the publicly funded research project "MEDIT" supported by the German Federal Ministry for Economic Affairs and Climate Action (BMWK) under grant number 0350028D.

connected to a bulk power system model (top-down approach) [11–14]. As manipulation scenarios, synchronized disconnection as well as active and reactive power setpoint adjustment are investigated. Though this method is suitable to depict the impact of the manipulated assets on the bulk power system itself, the local impact on the DGs remains unclear. Especially the propagation of cascading effects in the DGs through plant protection triggering and fault ride-through (FRT) characteristics of DERs (including the non-compromised assets) due to local voltage disruption cannot be assessed in detail with this top-down approach and the aggregated DG models.

Therefore, a detailed depiction of the DERs and DGs is necessary (bottom-up approach). The impact of malicious operation of DERs on the local bus voltages has thus far been investigated for single feeders [15, 16] or a medium voltage (MV) grid [17]. However, to investigate potential cascading effects, extended DG structures with high DER penetration have to be investigated. Hence, the focus of this paper is to simulate DGs with high DER penetration (WPPs, PVs) and their plant protection to investigate the impact of manipulated control signals sent to these DERs on stable power system operation. The contribution of the paper is as follows:

- Derivation of generalized manipulation scenarios and assessment criteria to quantify the criticality of manipulation scenarios targeting DERs
- Quantification of the additional impact of cascading effects caused by local voltage disruption
- Presentation of a case study to analyze the effect of increased DER penetration on the criticality of the manipulation strategy

## II. MODELING APPROACH

### A. Manipulation strategy and fundamental assumptions

A simplified schematic overview of the relevant parts of the CPS including the manipulation strategy is shown in Fig. 1.

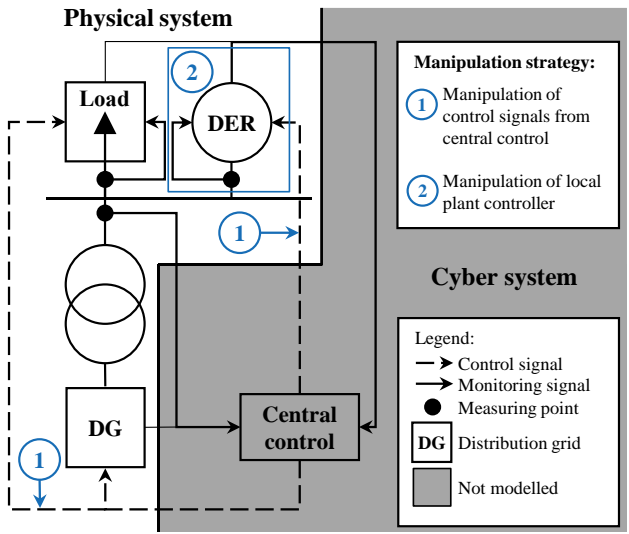

Figure 1: Overview of CPS

The assets of the power system (physical system) are connected to the central control system (cyber system) using ICT. TSOs and DSOs use supervisory control and data acquisition (SCADA) systems to monitor and control the distributed assets of the power system in real time (closed-loop control). For the communication between the central control system and the assets standardized protocols such as IEC 60870-5-104 or DNP3 are used. The communication is usually transmitted via exclusive communication channels but neither of these protocols support security mechanisms such as authentication or integrity protection [18]. Also, encryption is not used by a majority of SCADA system operators [19]. Although newer standards (e.g., IEC 62351) provide additional requirements for cyber security measures for each protocol, these are usually not implemented [18]. Therefore, the communication links are susceptible to cyber attacks. If the attacker get access, they can communicate with the assets within the specifications of the communication protocols. They can therefore manipulate control signals, forge additional ones or block any communication to and from the central control entity.

The following assumptions are made for the investigation:

1. All DERs at MV level and above are connected to a DSO control system via ICT. The assumption is valid, since this is already required in grid codes (e.g., [20] for German MV grids and [21] for German HV grids).

2. The attacker got access to the communication system of the DSO. In addition, he has sufficient knowledge of the control system, the power system and its assets due to previous long-term reconnaissance operations. Thus, he is able to deploy manipulation strategies via the communication system by sending control signals to all connected DERs using DI attacks. Past incidents (e.g., [6]) and their analysis show that this assumption is valid [22].

3. There is no immediate reaction from the DSO control system during the deployment of the manipulation strategy. Either the attacker is in direct control of the system or he impedes any data acquisition or reaction from it e.g., using DoS attacks [10]. Thus, no more closed-loop control between the DSO and the DERs exists.

4. The local plant control of the assets and the plant protection operate as designed. The control signals received are applied as defined in the grid codes. The plant control itself cannot distinguish whether the control command is compromised or not.

Using these assumptions, the modeling of the CPS is simplified. Due to assumptions 2 and 3, the cyber system containing the ICT system and the control system is not modeled. The power system and its components are modeled in detail. The component models contain an interface for the control signals from the control system and fulfil assumption 1. Assuming, that an attacker gained full control of the communication link and sufficient knowledge of all connected assets is a worst-case scenario. However, it is suitable to study and quantify the worst-case impact of the manipulation strategy on stable power system operation. It is comparable to classical transient stability analysis scenarios with their low probability but severe consequences (e.g., applying three-phase faults at different buses). The manipulation of the local plant controller of connected assets (manipulation strategy 2) offers additional degrees of freedom (e.g., override of plant protection). However, it is even less probable on a larger scale due to the high number of different vendors and thus not the focus of the investigations in this paper.

### B. Specified data points and grid code requirements

Grid codes are analyzed to find the data points transmitted between DERs and the central control. It is assumed, that all data points specified in the relevant grid codes (e.g., [20, 21] for German HV and MV grids) are transmitted. Monitoring signals (e.g., status, line voltage, and injected power) are sent from the assets to the control system of the grid operator as well as control signals (e.g., new setpoints for P and Q injection) in the opposite direction. Table I gives an excerpt of specified data points for generation units, storage units, and charging stations connected to the MV grid as required in [20].

TABLE I. SPECIFIED DATA POINTS (EXCERPT FROM [20])

| Signal | Type | Unit | Range |
|---|---|---|---|
| Line-earth-voltages $V_{LE}$ | Monitoring | kV | 0 - 15 |
| Line current $I_L$ | Monitoring | A | 0 - 2500 |
| Active power $P$ | Monitoring | kW | $\pm 120\% \cdot P_{inst}$ |
| Reactive Power $Q$ | Monitoring | kVAr | $\pm 50\% \cdot P_{inst}$ |
| Open tie breaker (emergency stop) | Control | - | Binary |
| Active power setpoint $P_{set}$ | Control | % | 0 - 100 |
| Reactive power setpoint $Q_{set}$ | Control | % | -50 - 50 |

In addition, the grid codes specify requirements for the DER control for set value application. New $P_{set}$ values from the grid operator shall be realized with a slope $\dot{P}$ defined as a percentage of installed P $P_{inst}$ of the DER per second as stated in (1) and defined in [20] and [21].

$$0.33 \tfrac{\%}{s} \cdot P_{inst} \leq |\dot{P}| \leq 0.66 \tfrac{\%}{s} \cdot P_{inst} \qquad (1)$$

These limits imply that the plants do not perform instantaneous changes for P injection after receiving a new P setpoint though it might be technically possible. The above-stated slope limits are not applicable if primary control is needed. For DERs connected at MV level, new $Q_{set}$ values shall be applied with a first-order lag behavior defined by the time constant $T$ as stated in (2) regardless of the chosen control mode for Q, i.e. Q(V), Q(P), fixed Q with V limitation, and fixed cos(φ) [20].

$$6 \text{ s} \leq 3 \cdot T \leq 60 \text{ s} \qquad (2)$$

Therefore, it takes at least 6 s for the Q injection to reach 95% of the new $Q_{set}$ value. If the value for the time constant $T$ is not specified by the DSO, a value of 10 s for the term $3 \cdot T$ is chosen. For DERs connected at HV level it is specified that the rise time from 0% to 90% $T_{0-90}$ is in between the range from 1 s to 60 s. If no value is specified by the grid operator $T_{0-90} = 5$ s is chosen [21].

### C. Simulation environment and component modeling

For the impact assessment of the manipulation scenario on stable power system operation, a Matlab-based symmetric RMS time-domain simulation is used. It contains generic dynamic models for generation units, loads, on-load tap changers (OLTCs) and their control circuits as well as plant protection functions (e.g., under-/overvoltage, reactive undervoltage, and under-/overfrequency protection). Additionally, low and high voltage ride through (LVRT/HVRT) characteristics are implemented as defined in the grid codes [23]. Detailed descriptions of the environment and its models are given in [24, 25]. The models are suitable for short-term stability analysis after large perturbations and can depict the relevant time constants of the control circuits, protection functions, and transient phenomena, that cannot be studied based on power flow calculations (i.e., the stationary operation point) [26]. The implemented models for the control circuits of the voltage source inverter (VSI) and the synchronous generator (SG) are parameterized to fulfill the requirements from the grid codes regarding new $P_{set}$ and $Q_{set}$ values as described in section II.B.

#### 1) Voltage source inverters (VSIs)

A generic VSI model is used for the representation of PV, full converter WPPs and battery storage systems (BSSs) [23]. The input values $I_{d,set}$ and $I_{q,set}$ for its current control loop are provided by the outer control loop. Additionally, a supplementary control is implemented to fulfil grid code requirements [25]. The outer control loop operates in P and Q control mode. The control loop for the P control is shown in Fig. 2 a).

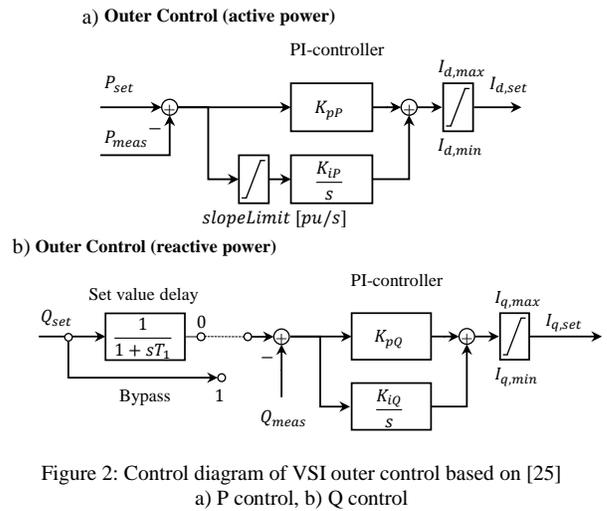

Figure 2: Control diagram of VSI outer control based on [25]
a) P control, b) Q control

A proportional integral (PI) controller is used to track the P setpoint by controlling the current $I_d$, which is limited to the maximum value $I_{d,max}$. A limiter for the P slope $\dot{P}$ is added to fulfill the requirements of P control stated in section II.B. The control loop for the Q also consists of a PI controller for the current $I_q$ but without the slope limit (see Fig. 2 b)). A first-order lag element with the time constant $T_1$ is added to enable the required behavior of the Q control. One additional limit is the maximum converter current limit $I_{max} = I_{d,max} = I_{q,max}$. If the injected current reaches this limit, the limitation prioritizes Q injection for grid support and reduces P injection accordingly.

*2) Synchronous generators (SGs)*

A sixth-order model of the SG is used to depict e.g., biogas plants in the DGs operating in grid parallel mode. Machine parameters are extracted from SG-datasheets [27]. The simplified governor model shown in Fig. 3 a) is used for P control [28]. By choosing the time constants $T$ and the gain $K$ this model can be used to approximate the dynamics for several governor types [28]. For P control in grid-parallel operation $K = 0$ can be used, thus neglecting droop control [29]. Therefore, the dynamic behavior is defined by the time constant $T_3$ and the limits. The requirements of Sec. II.B are also fulfilled by properly parameterizing the limiter for $\dot{P}_m$.

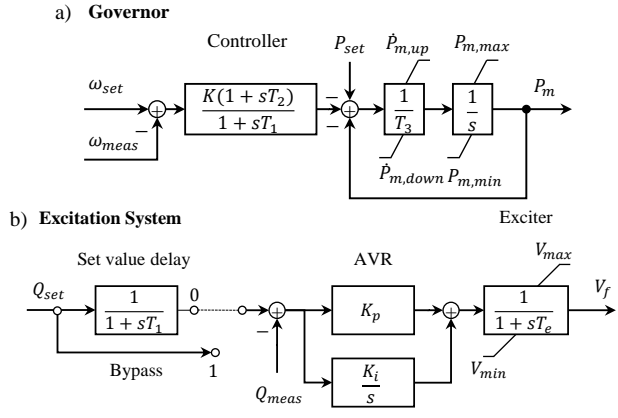

Figure 3: Control diagram of SG control
a) governor model based on [28], b) excitation system model

The generic excitation system model adapted from commercially available voltage regulators (AVRs) shown in Fig. 3 b) is used for Q control. The AVR is modeled as a PI controller and the exciter is modeled as a first-order lag element with a limiter neglecting saturation effects. The excitation voltage $V_f$ is the input for the SG model [24]. As for the VSI outer control, a first-order lag element is added to enable the required delayed application of new $Q_{set}$ values.

*3) On-load tap changers (OLTCs)*

The implemented discrete OLTC model based on [30] varies the ideal transformer's tap ratio $\tau$ to maintain the voltage $V_c$ at the primary or secondary winding within an acceptable range between the upper boundary $V_{ub}$ and lower boundary $V_{lb}$ of the defined voltage band as stated in (3) and (4).

$$\tau_{t+1} = \begin{cases} \tau_t + \Delta\tau & if\ V_c > V_{ub}\ \&\ trig = 1\ \&\ \tau_t < \tau_{max} \\ \tau_t - \Delta\tau & if\ V_c < V_{lb}\ \&\ trig = 1\ \&\ \tau_t > \tau_{min} \\ \tau_t & otherwise \end{cases} \quad (3)$$

$$V_{ub} = V_{ref} + db, V_{lb} = V_{ref} - db \quad (4)$$

The parameters of the tap changer are the time for performing a switching operation $T_s$, the tap ratio step $\Delta\tau$ and the maximum and minimum tap ratio $\tau_{max}$ and $\tau_{min}$. The trigger signal *trig* of the tap changer is set if $V_c$ remains outside the boundaries for a specified time delay $T_{tr}$. It can be defined using different time characteristics (e.g., constant, linear, and inverse). In addition, a two-stage sequential mode of operation can be depicted, where $T_{tr}$ varies depending if a first tap change has already occurred. If the trigger signal is set, the OLTC performs the corresponding switching operation and thus is blocked for the time $T_s$.

*D. Assessment criteria*

Power system frequency and bus voltages are suitable properties to assess the short-term impact of disturbances on power system operation. While power system frequency is a global property of the bulk power system mainly influenced by the P balance, bus voltages are a local property. For both properties, boundaries for normal operation are defined in standards (voltage/frequency band). If these values exceed their boundaries, plant protection functions can disconnect DERs. Therefore, the number and amount of P injection of disconnected DERs (including the amount of non-compromised DERs disconnecting due to local voltage band violations) is another property for assessing the disturbances. To assess these cascading effects in the DGs, a severity index based on deviation from voltage band and plant protection triggering as assessment criteria is designed (see Table II).

TABLE II. MANIPULATION SCENARIO SEVERITY INDEX

| Severity index | Deviation from voltage band | | Plant protection triggering | |
|---|---|---|---|---|
| | *During manipulation* | *After protection tripping* | *Manipulated DERs* | *Other DERs* |
| 0 | - | - | - | - |
| 1 | x | - | - | - |
| 2 | x | - | x | - |
| 3 | x & $d > 0.5$ | - | x | x |
| 4 | x & $d > 1$ | - | x | x |
| 5 | x & $d > 1$ | x | x | x |
| x: criterion fulfilled | | -: criterion not fulfilled | | |

Scenarios are rated from zero (non-critical) to five (extremely critical) based on the fulfillment of the assessment criteria. The maximum deviation $d$ from the voltage band is calculated using (5) with $X_{min/max}$ being the minimum/maximum measured bus voltage, $X_{boundary}$ the respective boundary, and $X_{ref}$ the reference value (in this case 1 pu for voltage):

$$d = \frac{|X_{boundary} - X_{min/max}|}{|X_{ref} - X_{boundary}|} \quad (5)$$

If e.g., the maximum voltage $V_{max}$ is 1.2 pu and the maximum voltage allowed by the voltage band $V_{boundary}$ is 1.15 pu the value for $d = 0.33$. The second assessment criterion is the plant protection triggering of DERs. If DERs are disconnected during the manipulation scenario due to their plant protection, the scenario is rated more critical. If the triggered DERs include non-compromised DERs that are not targeted by the manipulation scenario, the scenario is rated even more critical. The worst-case scenario is reached when even after the protection tripping the system remains outside the voltage band or is unstable. For MV level a voltage band of 0.9 pu to 1.1 pu and for HV level from 0.85 pu to 1.15 pu is used for $V_{boundary}$.

### III. CASE STUDY AND SIMULATION RESULTS

The modeling approach is applied to the following case study using two HV/MV benchmark grids expanded by dynamic component models and investigated using time-domain simulations. With the proposed method, the impact of the manipulation scenarios on stable powers system operation is quantified by applying the assessment criteria.

#### A. Investigated grid model and initial operating points

The investigated grid models consist of open source SimBench HV and MV benchmark grids for current and future scenarios (2024, 2034) [31]. All grid parameters are available and detailed explanation is provided in [32]. The predominantly rural HV1 grid with a rated voltage of 110 kV is characterized by a high share of overhead lines as well as isolated load [31]. In total, it consists of 64 stations connected by 95 lines with a total circuit length of 1,084 km [32]. The urban HV2 grid consists of 82 station connected by 113 lines with a total circuit length of 752 km. In comparison to HV1, a higher share of cabling and a higher level of intermeshing characterize it [32].

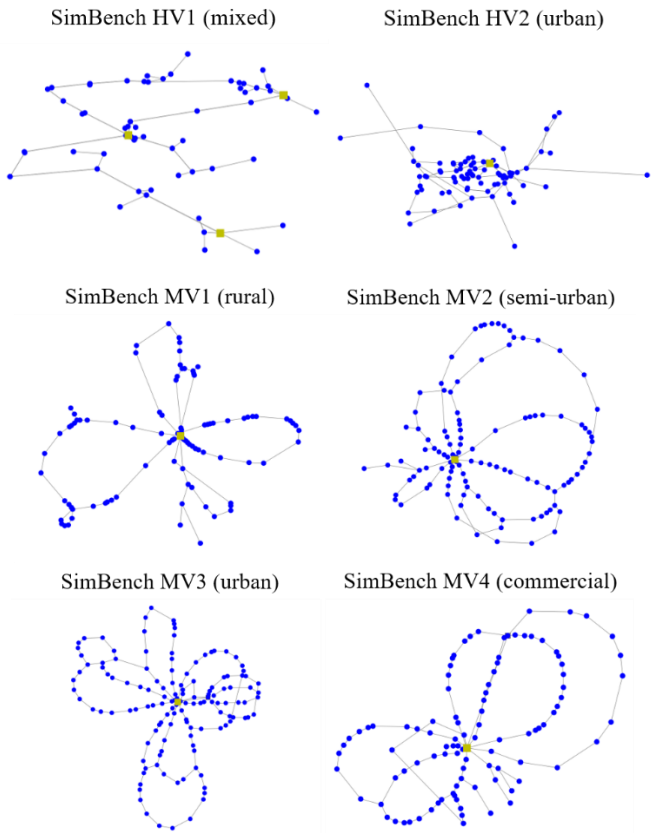

Figure 4: Topologies of SimBench HV and MV grid models

For the MV level (rated voltage 20 kV or 10 kV), SimBench provides four grid models (MV1-MV4) depicting rural, semi-urban, urban and commercial grids with their DERs and loads. Each low voltage subnetwork is modeled as one aggregated load and one aggregated DER (RES) [32]. The topologies of the HV and MV grids are shown in Fig. 4. It shows the buses (blue), branches (grey) and the grid connections points of the extra high voltage (EHV) level and the HV level, respectively (yellow). The two HV and the four MV grid models are combined to two HV-MV grid models with a current state and two future expansion scenarios (see Table A.I for grid model names). The bulk power system EHV level is modeled as one synchronous generator ($H$ = 6 s) providing the grid frequency and the necessary P and Q at each grid connection point to the HV level. The installed power of DERs for each grid model and expansion scenario (0-2) is shown in Table III.

TABLE III. SCENARIO SETUPS - INSTALLED POWER OF DERS

| Grid model | Wind HV (MW) | Wind MV (MW) | PV HV (MW) | PV MV (MW) | Hydro MV (MW) | Biomass MV (MW) | RES LV (MW) |
|---|---|---|---|---|---|---|---|
| HV-MV-mixed-0 | 1077.2 | 126.1 | - | 9.3 | 15.2 | 9.2 | 173.4 |
| HV-MV-mixed-1 | 1295.0 | 216.2 | - | 9.3 | 15.2 | 9.2 | 328.1 |
| HV-MV-mixed-2 | 1932.2 | 254.2 | - | 9.3 | 15.2 | 9.2 | 391.6 |
| HV-MV-urban-0 | 300.0 | 107.1 | 10 | 11.4 | 12.4 | 6.5 | 138.8 |
| HV-MV-urban-1 | 451.4 | 207.4 | 106.6 | 11.4 | 12.4 | 6.5 | 252.1 |
| HV-MV-urban-2 | 1167.5 | 240.4 | 276.4 | 11.4 | 12.4 | 6.5 | 310.9 |

Dynamic parameters for generation units (SGs, VSIs), their control circuits, plant protection and OLTCs are added to perform time-domain simulations (see Tables A.II and A.III for VSI and OLTC parameters). A constant power ZIP load model is used for all loads. The storage systems in the future expansion scenarios are deactivated and do not contribute. For the stability analyses, five study cases from the SimBench dataset are chosen ($hL$, $hW$, $hPV$, $lW$, $lPV$). They define the initial operating points for the DERs and loads as well as slack bus voltages and reference voltages for the OLTCs. They represent the boundary values of initial operating points from low load and high DER generation to high load and low DER generation (see Table A.IV for detailed description). They also include a distinction between Wind and PV generation [32]. Although originally defined for planning studies, they prove to be well suited for stability analyses to account for the diversity of operating points of the power system and its assets in a simplified manner. For each study case the OLTCs are initialized (first EHV/HV OLTCs then HV/MV OLTCs) by adjusting their tap position until the initial bus voltages at the controlled buses are within the OLTC deadbands defined in Table A.III.

#### B. Investigated manipulation scenarios

In this study we consider the worst-case, i.e. an attacker can send manipulated setpoints to the DERs with a guaranteed success. Five exemplary manipulation scenarios are derived based on the accessible control signals for P and Q setpoints as well as the emergency disconnection command (see Table IV).

TABLE IV. MANIPULATION SCENARIOS

| Manip. Scenario | Time (s) | Applied to | Variant 1 | | Variant 2 | |
|---|---|---|---|---|---|---|
| | | | $P_{set}$ (pu) | $Q_{set}$ (pu) | $P_{set}$ (pu) | $Q_{set}$ (pu) |
| Undervoltage | 0.5 | WPPs | 0 | - | 0 | - |
| | 1 | WPPs | - | -0.33 | - | -0.5 |
| Overvoltage | 1 | WPPs | - | 0.33 | - | 0.5 |
| Disconnection | 1 | WPPs | - | | - | |

The target of the undervoltage (*uv*) and overvoltage (*ov*) manipulation scenarios is to utilize the P and Q supply of all WPPs at HV and MV level to force the bus voltages outside the voltage bands. During the scenarios *uv1* and *uv2* the WPPs receive new setpoints for $P_{set}$ at t = 0.5 s and $Q_{set}$ at t = 1 s. The decrease of P injection as well as the increased Q consumption lead to decreasing bus voltages at the grid connection points of the WPPs. Thus, e.g., undervoltage protection functions are triggered. For the scenarios *ov1* and *ov2* new setpoints $Q_{set}$ are sent to the WPPs at t = 1 s to increase their Q injection. No new setpoint $P_{set}$ is sent, since it is assumed that the WPPs are already operating at maximum P injection defined by the study case. In variant *1* $Q_{set}$ is limited to ±0.33 pu as it is specified in [20]. In variant *2* it is assumed, that the converter of the WPP is able to provide Q outside the limits of [20] up to the maximum range of the control signal as specified in Table I. Thus $Q_{set}$ = ±0.5 pu is chosen. For the disconnection manipulation scenario (*disc*) all WPPs at HV and MV level are disconnected simultaneously at t = 1 s. Therefore, the P supply of the WPPs as defined in the study cases is immediately lost. Each manipulation scenario is simulated for $t_{sim}$ = 300 s to include the time constants of P and Q control, tap changer operation as well as plant protection functions.

*C. Overview of results*

In Fig. 5 the severity index of all 150 scenarios of the case study is presented for each combination of grid model, study case and manipulation scenario. The 31 non-critical scenarios with a severity index of *0* are mainly current grid scenarios. For these scenarios, no voltage band violations and thus no protection triggering can be achieved with the manipulation scenarios. The local impact in the DGs of these manipulation scenarios is therefore non-critical. In addition, there is also no frequency band violation for the synchronous disconnection of the WPPs in the *disc* scenarios. The overall severity increases for the future expansion scenarios of the grid models with increased DER (and thus WPP) penetration. There is more installed power of WPPs available that can be utilized by the attacker during the manipulation scenarios. Thus, the influence on the bus voltages of the DGs is higher. Therefore the manipulation scenarios lead to a higher criticality in the DGs indicated by the higher severity index. As a consequence of the manipulation scenarios, voltage band violations (severity index *1*) as well as plant protection triggering of the targeted WPPs (severity index *2*) and also other DERs (severity index *3* to *5*) occurs. For scenarios with a severity index of *4* and *5*, the maximum deviation *d* from the voltage band is larger than 1. One scenario is identified as extremely critical with a severity index of *5*. Even after plant protection triggering and OLTC operation the bus voltage remain outside the voltage band due to manipulated WPPs still absorbing reactive power and not being disconnected by undervoltage or reactive undervoltage protection. The manipulation scenarios *ov1* and *ov2* have an increased severity for the low load and high generation study cases *lPV* and *lW* due to their higher initial bus voltages (smaller margin to upper voltage band limit), while the manipulation scenarios *uv1* and *uv2* affect mostly the high load study cases *hL*, *hPV* and *hW* with generally lower initial bus voltages (smaller margin to lower voltage band limit). If the manipulated assets can only provide limited Q of ±0.33 pu as assumed in *ov1* and *uv1* the severity of the manipulation scenario is lower compared to *ov2* and *uv2*, where Q injection and consumption is only limited by the range of the control signal values ($Q_{set}$ = ±0.5 pu). The *disc* scenarios are either non-critical in current grid scenarios or have a severity index of *2* in future expansion scenarios. Their ability to violate bus voltages and force protection triggering of other assets is limited, since all WPPs are immediately disconnected and the OLTCs can bring back bus voltages within the voltage band quickly.

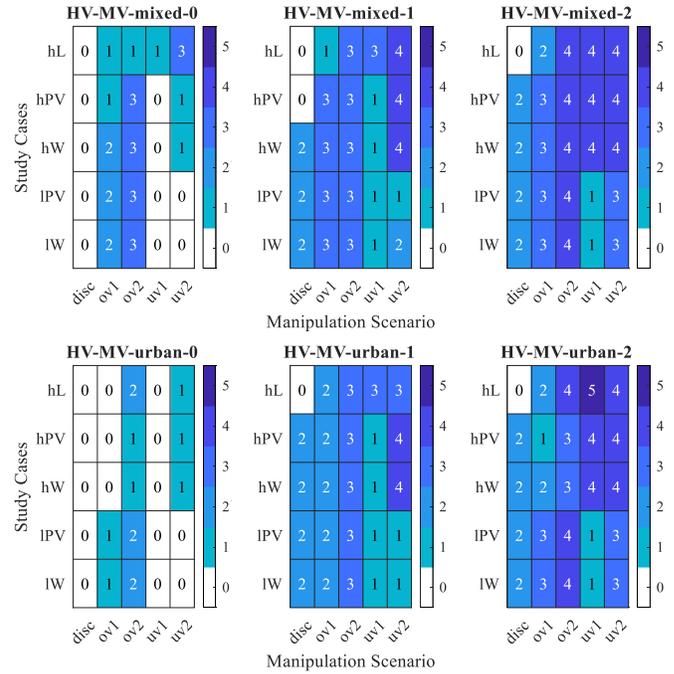

Figure 5: Severity index for all scenarios of case study

The metrics lost P injection and number of triggered plant protection units for each scenario are shown in Fig. 6. The scenarios are divided into *disc* scenarios (left) and *ov/uv* scenarios (right). The scenarios of each category are sorted descending by lost P injection in both plots. Four exemplary scenarios are highlighted, which are discussed in the following sections. For clarity, the *ov/uv* scenarios with no lost P injection are not shown. The share of each metric, which corresponds to plants targeted by the attacker (i.e., the WPPs), is indicated in orange while the share corresponding to additional plants, which were not targeted by the attacker, is indicated in blue. The *disc* manipulation scenarios *HV-MV-mixed-2-lW* and *HV-MV-mixed-2-hW* lead to the highest amount of lost P injection since all WPPs operating at maximum P injection of 2,186 MW are disconnected. For the *ov/uv* scenarios, a maximum amount of 785.2 MW is disconnected in scenario *HV-MV-mixed-2-hW-uv2* with 619.5 MW belonging to targeted WPPs. In total 733 plants are disconnected (including aggregated LV RES plants) with 37 being part of the manipulated WPPs.

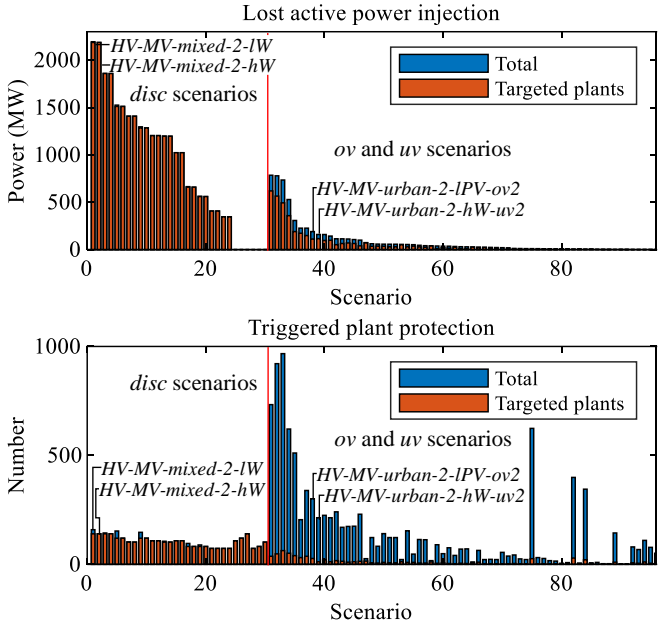

Figure 6: Lost P injection and number of triggered plant protection

Although the the amount of lost P injection is lower for the *ov/uv* scenarios, the number of disconnected plants due to plant protection triggering is significantly higher (with a maximum of 967 DERs being disconnected) compared to the *disc* scenarios. In these scenarios, the manipulated WPPs can trip a significant amount of neighboring DERs by influencing their bus voltages mainly by injecting or absorbing Q so that the limits of the voltage band are exceeded long enough.

### D. Analysis of exemplary scenarios

The time characteristics of the substation bus voltages, OLTC tap positions and the power supply from the EHV grid are shown in Fig. 7 for three exemplary scenarios. These scenarios are further discussed in the following sections.

#### 1) Scenario I

In Fig. 7 a) the scenario *HV-MV-urban-0-hW-ov2* with a severity index of *1* is shown. The study case *hW* specifies a high load and high WPP infeed scenario. The target of the overvoltage manipulation scenario *ov2* is to force the bus voltage above the upper voltage limit by injecting Q of 0.5 pu by all WPPs. Hence, the bus voltages at nearby HV and MV buses increase (only substation buses are shown). Since the bus voltage at HV substation bus 1391 with a value of 1.041 pu is outside the OLTC deadband of the connected EHV/HV transformers (upper limit of 1.04 pu), the tap changers adjust their position once at t = 33 s (new tap position of 5). After this adjustment, the bus voltages at the HV and MV buses decrease and the bus voltage at HV bus 1391 with a value of 1.032 pu is back within the OLTC deadband. However, the MV bus voltages at several HV/MV substations are outside the OTLC deadbands of the HV/MV transformers (even above the MV voltage limit of 1.1 pu at MV bus 471). Therefore, tap changers at five HV/MV substations start adjusting their tap positon and the voltage at MV bus 471 returns into the voltage band. Due to the fast OLTC operation and high load study case *hW* the impact of Q injection is not sufficient to increase bus voltages at any bus above the upper voltage limit for longer than 60 s and subsequently trigger overvoltage plant protection. At the grid connection point to the EHV level, Q flow is changed by 197 MVAr in the first 15 s, while the P flow remains at the same level. From the perspective of the bulk power system, the impact of this specific scenario is thus negligible. However, the manipulated WPPs injecting reactive power remain connected to the DG and thus to the bulk power system.

#### 2) Scenario II

In contrast, the scenario *HV-MV-urban-2-hW-uv2* with a severity index of *4* is shown in Fig. 7 b). With all WPPs decreasing their P infeed to zero and absorbing Q of 0.5 pu, the bus voltages at several HV and MV buses are quickly forced below the lower limit of the voltage band during the first 15 s. Therefore, after 0.5 s Q(V)-protection functions of two HV and seven MV WPPs trigger since the bus voltage at the plants is below 0.85 pu and disconnect 97.73 MW of P infeed during the first 7 s. Three additional manipulated WPPs disconnect after fulfilling LVRT specifications adding 18.1 MW to the lost P infeed. Though the loss of P infeed generally leads to further decreasing bus voltages, the end of Q absorption of the manipulated plants compensates the effect. Besides the manipulated WPPs, 43.9 MW of aggregated LV DERs disconnect after remaining connected as specified by LVRT requirements. A steady state is reached until the beginning of OLTC operation after their specified time delays of 25 s (EHV/HV) and 55 s (HV/MV) at t = 27 s. The EHV/HV and HV/MV OLTCs are able to return their controlled bus voltages into the respective voltage deadband. This process leads to all bus voltages at the substations to get back into the voltage band within the first 100 s. However, 86 manipulated WPPs remain active and continue to reduce their P infeed to zero. This necessitates further tap changer operation until t = 200 s. Afterwards a steady state is reached with all voltages back within the voltage band. If this state kept, this will prevent further DERs from disconnecting due to undervoltage or overvoltage protection for the following 60 minutes and the immediate voltage collapse is prevented. However, the significant loss in generation (1432 MW within 150 s) as well as the change in Q supply (500 MW within 150 s) have to be compensated by the bulk power system using its control power.

#### 3) Scenario III

The third scenario *HV-MV-urban-2-lPV-ov2* with a severity index of *4* is shown in Fig. 7 c). With a low load initial operating point ($lPV$) and each WPP injecting Q of 0.5 pu, the HV and MV bus voltages increase above the upper voltage limit of 1.15 pu at the HV substation buses and 1.1 pu at the MV substation buses. This leads to the disconnection of 278 MV DERs with a total lost P injection of 142.9 MW after fulfilling HVRT requirements (stay connected for bus voltage above 1.2 pu for 0.1 s) during the first 25 s. Though OLTC operation starts, it is not fast enough to bring back voltages into the voltage band at all HV and MV buses within 60 s. Therefore, after 60 s overvoltage protection disconnects additional 47.4 MW of HV and MV plants since the bus voltages are still above 1.15 pu and 1.1 pu respectively. While 27 manipulated WPPs are disconnected by plant protection, 99 manipulated WPPs remain active and inject P and Q. However, the manipulation scenario will achieve no further protection triggering. For the bulk power system the P flow at the grid connection point changes within the range of the lost P injection of DERs, while Q flow changes by 698 MVAr within the first 70 s.

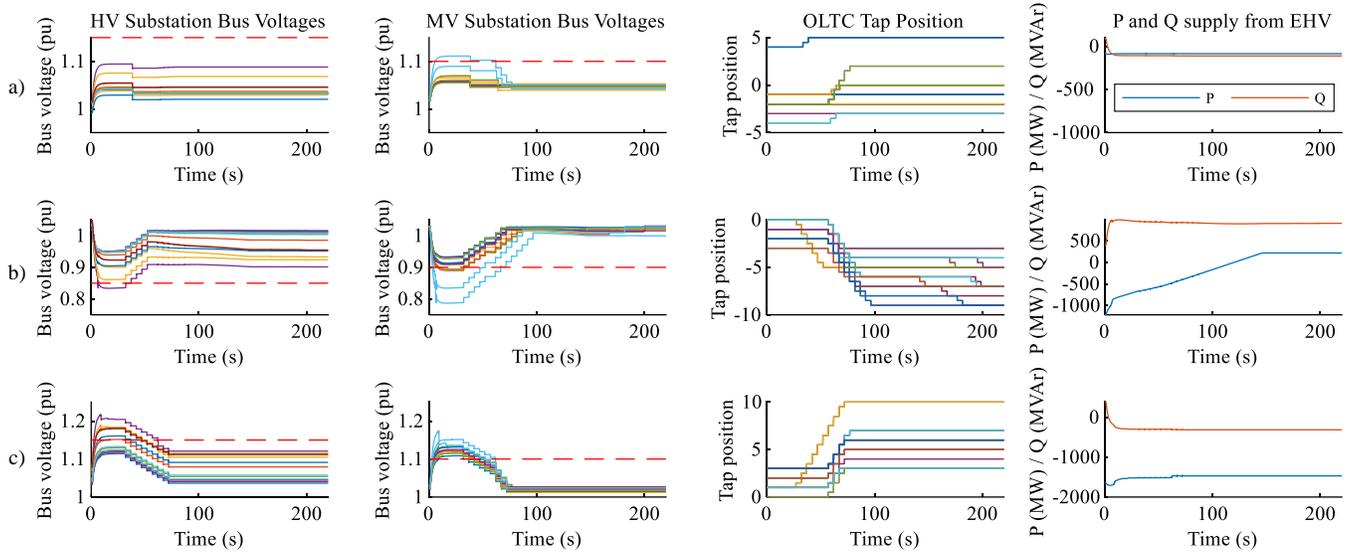

Figure 7: HV and MV substation bus voltages and OLTC tap position for exemplary scenarios: a) Scenario I, b) Scenario II, c) Scenario III

*E. Discussion*

Based on the obtained results it is clear, that the disconnection commands as well as the manipulated setpoints for WPPs and the subsequent protection triggering lead to significant short-term changes in the operating point of the DGs within the first 30 s. To compensate for these changes the bulk power system must provide or absorb P and Q within this time scale, since all DERs in the DGs are assumed to have fixed PQ setpoints and loads are assumed as constant power loads. By modeling the bulk power system as large synchronous generators without dedicated voltage/frequency control for this case study it is implicitly assumed that the bulk power system can provide or absorb the necessary P and Q without significant impact on system frequency or EHV bus voltages. Though the disconnection of DERs is straight forward and leads to the highest amount of lost P injection, the manipulation of Q setpoints offers a method to change the local bus voltages at HV and MV buses significantly and achieve cascading effects. Since the DERs connected to the HV and MV grid have to fulfill the grid code requirements that define the speed of the setpoint application (usually within 15 s) this change is fast, especially compared to the application of P setpoints. In general, increased DER penetration (e.g., in the future expansion scenarios *1* and *2*) leads to more critical manipulation scenarios, with the worst-case assumption, that the attacker the attacker can also manipulate the control signals of the additional DERs.

If the communication and the central control loop between DERs and central control are compromised, local control loops in the DGs proof as suitable countermeasures. Automated and reasonable fast local control of OLTCs at EHV/HV and HV/MV substations can prevent triggering of over-/undervoltage protection of additional plants. It mitigates the impact of the manipulation scenarios if bus voltages are brought back into the voltage band within 60 s before additional over-/undervoltage protection tripping of the DERs. Other voltage control devices –though not investigated in this case study– are also a suitable option. Also, the limitation of the Q range that can be accessed by remote control signals proves beneficial as can be seen by the comparison of the criticality between the variant *1* and *2* of the *ov* and *uv* manipulation scenarios. Therefore, it is recommended to limit the range of the Q setpoint to a range that is sufficient for stable grid operation e.g., to ±0.33 pu or further, or to further limit absorption/injection of Q by DERs during undervoltage/overvoltage situations. It has to be noticed that since manipulated assets always influence their own bus voltage with their P and Q injection their plant protection often disconnect them too (see Fig. 6). Though this contributes to further lost P injection, it also reduces the number of potential assets available for the attacker if it is assumed that the plant protection triggering cannot be reset by the attacker.

This investigation focuses on the impact and possible countermeasures in the power system. However, grid operators must implement suitable cyber security measures to prevent or impede this manipulation scenario (e.g., encryption of communication between control and DERs) as well as incident response strategies to regain control of the ICT system and prevent further action by the attacker (e.g., sending additional control commands).

## IV. Conclusion and Outlook

This research work assesses the impact of manipulated DERs on stable power system operation. A special focus is set on quantifying local cascading effects in the DGs caused by DER control and protection functions. Using generic manipulation scenarios, time-domain simulations and a severity index based on voltage band violation and plant protection triggering, a quantification of the impact is achieved. This is demonstrated in a case study using HV-MV benchmark grids with different study cases as initial operating points for DERs and loads. As manipulation strategy, compromised control commands for disconnection as well as P and Q setpoints are sent to all WPPs in the benchmark grids to achieve voltage band violations and to trigger subsequent plant protection actions. Though the amount of lost P is higher in the *disc* scenarios, the manipulation of P and Q setpoints in the *ov/uv* scenarios can significantly change bus voltages locally and thus trigger additional plant protection of other assets. Furthermore, high DER penetration as well as a high amount of accessible Q increase the severity of the *ov/uv* scenarios. Cyber attacks manipulating DERs in the DGs can cause severe consequences for the bulk power system, if done on a sufficiently large scale. On the other hand, automated OLTCs at EHV/HV and HV/MV substations have a mitigating effect if they act fast enough to bring the bus voltages back into the voltage band before additional plant protection triggering (e.g., over-/undervoltage protection disconnection after 60 s). Also, a limitation of the remotely accessible Q injection or local DER control prohibiting Q injection and absorption depending on the bus voltage situation can prove beneficial.

In future work, the impact of manipulation scenarios targeting bulk power system frequency control and the effect of varying bulk power system inertia need to be evaluated. Also, the impact of the manipulation of other assets (e.g., controllable loads such as charging stations) can be quantified. Additional analysis can address situations, where the success of the attacker sending setpoints to the DERs is not ensured, i.e. likelihood of less than one.

APPENDIX

TABLE A.I.    INVESTIGATED GRID MODELS

| SimBench codes of investigated grid models | |
|---|---|
| 1-HVMV-mixed-all-0-no_sw | 1-HVMV-urban-all-0-no_sw |
| 1-HVMV-mixed-all-1-no_sw | 1-HVMV-urban-all-1-no_sw |
| 1-HVMV-mixed-all-2-no_sw | 1-HVMV-urban-all-2-no_sw |

TABLE A.II.    VSI CONTROL PARAMETERS FOR WPP & PV

| Parameter | WPP | PV |
|---|---|---|
| Time delay current control $T_d$, $T_q$ | 10 μs | 10 μs |
| Maximum current $I_{max}$ | 1.1 pu | 1.1 pu |
| Integral gain outer control P $K_{iP}$ | 50 | - |
| Proportional gain outer control P $K_{pP}$ | 0.001 | - |
| Integral gain outer control Q $K_{iQ}$ | 50 | - |
| Proportional gain outer control Q $K_{pQ}$ | 0.001 | - |
| Set value delay Q control $T_1$ (MV) | 3.33 s | - |
| Set value delay Q control $T_1$ (HV) | 2.17 s | - |
| Slope limit for P control | 0.0066 pu | - |
| Integral gain outer control AC $K_{iAC}$ | - | 0 |
| Proportional gain outer control AC $K_{pAC}$ | - | 2 |
| Integral gain outer control DC $K_{iDC}$ | - | 166.67 |
| Proportional gain outer control DC $K_{pDC}$ | - | 5 |
| Converter's capacity $C_{DC}$ | - | 0.172 s |

TABLE A.III.    OLTC PARAMETERS

| Parameter | OLTC EHV/HV | OLTC HV/MV |
|---|---|---|
| Switching time $T_s$ | 5 s | 5 s |
| Tap ratio Δτ | 0.01 pu | 0.01 pu |
| Maximum tap position | 16 | 9 |
| Minimum tap position | -16 | -9 |
| Half of OLTC deadband db | 1.5% | 2% |
| Reference voltage $V_{ref}$ (hL, hW, hPV) | 1.025 pu (HV) | 1.035 pu (MV) |
| Reference voltage $V_{ref}$ (lW, lPV) | 1.025 pu (HV) | 1.015 pu (MV) |
| First step constant time delay $T_{tr1}$ | 25 s | 55 s |
| Consecutive time delay $T_{tr}$ | 5 s | 5 s |

TABLE A.IV.    SCALING FACTROS FOR THE DEFINED STUDY CASES [32]

| Study case | Description | Load | | Generation | | |
|---|---|---|---|---|---|---|
| | | | | Wind | PV | Other |
| | | p | cos(φ) | p | p | p |
| hL | high load, no DER generation | 1 | 0.93 | 0 | 0 | 0 |
| hW | high load, very high Wind, high PV, high other DERs | 1 | 0.93 | 1 | 0.8 | 1 |
| hPV | high load, high Wind, very high PV, high other DERs | 1 | 0.93 | 0.85 | 0.95 | 1 |
| lW | low load, very high wind, high PV, high other DERs | 0.25 (HV), 0.1 (MV) | 0.9 | 1 | 0.8 | 1 |
| lPV | low load, high wind, very high PV, high other DERs | 0.25 (HV), 0.1 (MV) | 0.9 | 0.85 | 0.95 | 1 |